# Localisation of hexagonal boron nitride colour centres using patterned dielectric layers on graphene

M. K. Prasad,[1, 2] V. Babenko,[3] A.W. Tadbier,[3] S. Hofmann,[3] J. P. Goss,[1] and J. D. Mar[1, 2]

[1)]*School of Mathematics, Statistics and Physics, Newcastle University, Newcastle upon Tyne, NE1 7RU, United Kingdom*

[2)]*Joint Quantum Centre (JQC) Durham-Newcastle, United Kingdom*

[3)]*Department of Engineering, University of Cambridge, Cambridge, CB3 0FA, United Kingdom*

(*Electronic mail: jonathan.mar@newcastle.ac.uk)

(*Electronic mail: sh315@cam.ac.uk)

(Dated: 12 March 2025)

One of the most promising building blocks for the development of spin qubits, single-photon sources, and quantum sensors at room temperature, as well as 2D ultraviolet light-emitting diodes, are defect colour centres in 2D hexagonal boron nitride (hBN). However, a significant requirement for the realisation of such devices towards scalable technologies is the deterministic localisation of hBN colour centres. Here, we demonstrate a novel approach to the localisation of hBN colour centre emission by using patterned dielectric layers grown on graphene via atomic layer deposition (ALD). While colour centre emission is quenched within areas where hBN is deposited directly on graphene due to charge transfer, it is maintained where hBN is deposited on micron-sized $Al_2O_3$ pillars which act as barriers to charge transfer. Importantly, our approach allows for device architectures where graphene layers are used as top and bottom electrodes for the application of vertical electric fields, such as for carrier injection in electroluminescent devices, Stark shifting of colour centre emission, and charge state control of defect colour centres. Furthermore, the use of ALD to grow dielectric layers directly on graphene allows for the control of tunnel barrier thicknesses with atomic layer precision, which is crucial for many of these device applications. Our work represents an important step towards the realisation of a wide range of scalable device applications based on the determinsitic localisation of electrically controlled hBN colour centres.

Defect colour centres in 2D hexagonal boron nitride (hBN) have gained significant interest recently for the realisation of spin qubits[1,2], single-photon sources[3], and quantum sensors[4] at room temperature, as well as 2D ultraviolet light-emitting diodes (LEDs)[5–7]. A major challenge in the development of such devices into scalable technologies is the localisation of hBN colour centres into well-ordered arrays. While the natural complementarity between hBN and graphene in hBN/graphene heterostructures has led to the creation of a wide variety of novel 2D device applications[5–10], previous work[11] has demonstrated the localisation of hBN colour centre emission within windows patterned in graphene substrates where the quenching of colour centre emission due to charge transfer to graphene[12,13] was avoided. However, such an approach to localisation is incompatible with device architectures where graphene layers are used as electrodes for the application of vertical electric fields perpendicular to the hBN plane, such as for carrier injection in electroluminescent devices[5–7,9], Stark shifting of colour centre emission[14], and charge state control of defect colour centres[15].

In this Letter, we demonstrate a novel approach to the localisation of hBN colour centre emission by using patterned dielectric layers grown on graphene via atomic layer deposition (ALD). While colour centre emission is quenched within areas where hBN is deposited directly on graphene due to charge transfer[12,13], it is maintained where hBN is deposited on the dielectric layer which acts as a barrier to charge transfer. Importantly, unlike previous work[11], our approach to the localisation of hBN colour centre emission allows for device architectures where graphene layers are used as top and bottom electrodes for the application of vertical electric fields[5,7,9,14,15]. Moreover, unlike previous works[5,7,9,14,15], our ability to use ALD to grow dielectric layers directly on graphene allows for the control of tunnel barrier thicknesses with atomic layer precision, which is especially important for devices with applications in electroluminescence[5–7,9] and charge state control of colour centres[15]. The results of this work will therefore open the way for a wide variety of scalable device applications through the realisation of well-ordered arrays of electrically controlled hBN colour centres.

The device structure used in this work to achieve localisation of hBN colour centre emission is shown schematically in Fig. 1. Large-area and continuous monolayer graphene is de-posited on a $SiO_2$/Si substrate. Following the growth of a thin dielectric layer of $Al_2O_3$ directly on the graphene layer via ALD, the $Al_2O_3$ layer is then patterned into arrays of micron-sized square pillars via electron-beam lithography (EBL). Finally, hBN flakes with a lateral size of 50–200 nm and a thickness of 1–5 monolayers are then deposited over the entire structure. While colour centre emission from hBN flakes deposited directly on graphene is quenched due to charge transfer, it is maintained for hBN flakes deposited on $Al_2O_3$ pillars due to $Al_2O_3$ serving as a barrier to charge transfer, in addition to serving as a spacer layer to reduce spectral diffusion[16]. As a result, hBN colour centre emission is localised deterministically to the lithographically-defined $Al_2O_3$ pillars.

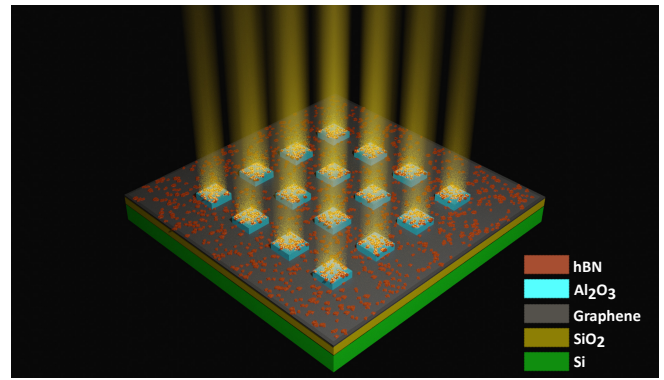

FIG. 1. Schematic of the device structure for the localisation of colour centre emission from hBN flakes using an $Al_2O_3$ dielectric layer grown directly on graphene via ALD and patterned into an array of micron-sized square pillars via electron-beam lithography.

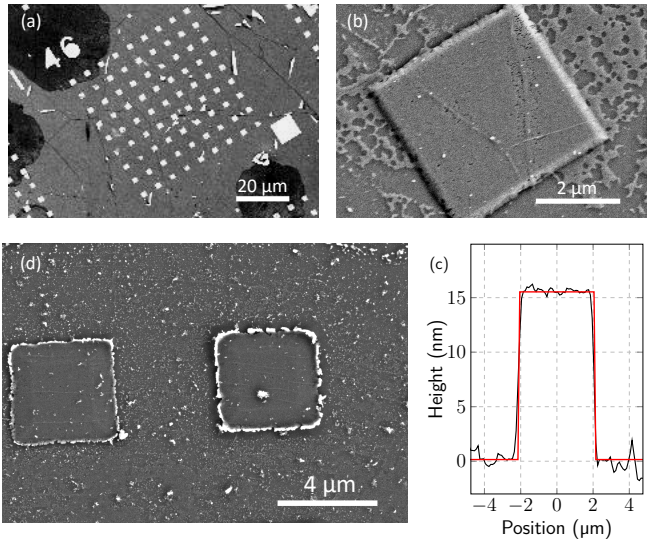

FIG. 2. **(a)** SEM image of a $9 \times 9$ array of $Al_2O_3$ square pillars on graphene, obtained via EBL and wet etching. **(b)** High-magnification SEM image of an individual $Al_2O_3$ pillar on graphene. **(c)** AFM measurement of the step height of the pillar in **(b)**, where the solid red line is a fit using the Gwyddion software[17]. No postprocessing was performed since the measurement background was negligible. **(d)** SEM image of hBN flakes deposited on the sample. Flakes tend to accumulate towards the edges of pillars due to a 'coffee-ring' effect.

We now describe the details of the fabrication process for our device shown schematically in Fig. 1. Large-area and continuous monolayer graphene transferred onto $SiO_2$/Si substrates using a polymer-free process were obtained from a commercial supplier (Grolltex), since graphene transfer processes involving polymers may lead to the presence of polymer residues where the preferential nucleation of precursors would result in a non-uniform ALD layer. While it is well known that the inertness of graphene presents a challenge for the growth of $Al_2O_3$ layers on graphene via ALD[18], it has been found that prolonged 'soaking times' of precursors was able to compensate for the slow reaction kinetics[18]. Following this recipe, we were therefore able to perform ALD at a high temperature of 200°C while avoiding desorption of nucleation sites from the graphene surface. High deposition temperatures offer the advantage of increasing the dielectric constant of deposited layers, which in turn further suppresses quench-ing of colour centre emission through Förster resonance energy transfer (FRET)[19]. By performing ALD of $Al_2O_3$ on graphene for various numbers of cycles ranging from 100 to 500 cycles, we determined the growth rate to be 0.13 nm per cycle, as confirmed by ellipsometry of the layer thickness. While previous studies have shown that a 50-nm-thick $Si_3N_4$ barrier between CdSe/ZnS core-shell quantum dots and graphene[20] and a 20-nm-thick $SiO_2$ barrier between fluorescent dyes and graphene[21] were sufficient to suppress quenching of emission and given that these barrier materials have comparable dielectric constants to $Al_2O_3$, we decided to use a 40-nm-thick $Al_2O_3$ barrier between hBN and graphene in our device in order to suppress quenching of hBN colour centre emission.

We then patterned the $Al_2O_3$ layer into arrays of square pillars via EBL and wet etching in a 3:1 solution of $H_3PO_4$:DI water at a temperature of 80°C for 210 s. Since the precise density of colour centre emitters cannot be known prior to the deposition of hBN flakes, the arrays of square pillars were patterned with lateral dimensions ranging from $0.5 \mu m \times 0.5 \mu m$ to $4.75 \mu m \times 4.75 \mu m$, allowing for the possibility to obtain pillars with a dimension that results in the localisation of approximately one colour centre emitter per pillar. Figure 2a shows a low-magnification scanning electron microscopy (SEM) image of a $9 \times 9$ array of $Al_2O_3$ pillars. It is important to notice that several multilayer pyramids and wrinkles of the underlying graphene are still present, confirming that the graphene layer has been preserved following wet etching of the $Al_2O_3$ layer. Using an Everhart-Thornley detector for enhanced SEM imaging of the pillar surface, Fig. 2b shows a high-magnification SEM image of an individual $Al_2O_3$ pillar obtained from 100 ALD growth cycles and patterning into a $4.75 \mu m \times 4.75 \mu m$ square pillar via EBL and wet etching. Successful etching of $Al_2O_3$ down to the graphene layer in order to form the pillar with the desired step height was confirmed by measuring the step profile across the pillar via atomic force microscopy (AFM). As shown in Fig. 2c, the step height was measured to be $15.37 \pm 0.23$ nm, in agreement with the predicted step height based on the established ALD growth rate. It can also be seen in Fig. 2b that, although most of the $Al_2O_3$ away from the pillar has been removed, residual islands of $Al_2O_3$ still remain. These islands are the result of the polycrystalline nature of the ALD film, leading to a non-uniform etch rate of the $Al_2O_3$ layer. Therefore, the etch time was optimised to ensure maximal removal of the $Al_2O_3$ layer away from the pillars, while also minimising undercutting of the $Al_2O_3$ pillars (as shown in Fig. 2c). As the thickness of these residual islands was confirmed by AFM to be less than 5 nm, they should not prevent quenching of hBN colour centre emission away from the pillars[20,21].

Having patterned the arrays of $Al_2O_3$ pillars on graphene, the hBN flakes in a 50:50 ethanol:water solution (Graphene Supermarket) were sonicated for 30 min, dropcast over the entire sample, and left to dry. As the drying process sometimes leads to self-rolling of the graphene layer to form nanoscrolls[23,24], a thin uniform layer of $Al_2O_3$ (∼5 nm) was grown via ALD over the entire sample prior to dropcasting hBN flakes, serving as a protective layer to prevent the self-rolling of graphene. From an SEM image of hBN flakes deposited on the sample (Fig. 2d), it can be seen that flakes tend to accumulate towards the edges of the pillars. This may be due to a 'coffee-ring' effect which has been previously observed following the evaporation of nanosphere colloidal drops on the top of pillars[25].

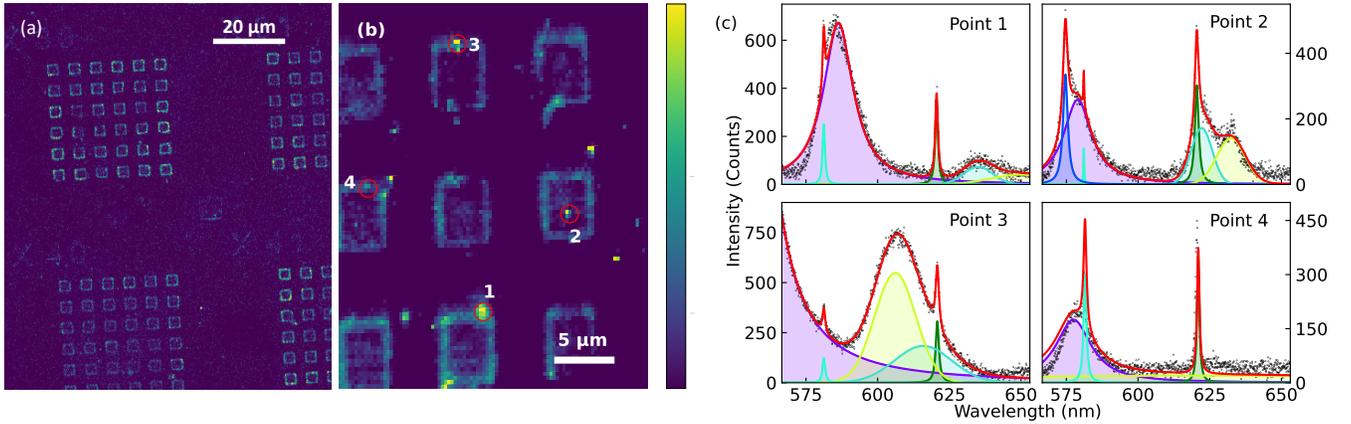

FIG. 3. **(a)** Confocal PL map of a sample area where hBN flakes have been deposited and where 6 × 6 arrays of 40-nm-thick 4.75 $\mu$m × 4.75 $\mu$m Al$_2$O$_3$ pillars have been patterned on graphene. The laser excitation wavelength was 473 nm and the detection wavelength range was 560–660 nm. It is clear that emission is localised at the pillars. **(b)** Hyperspectral confocal PL map of hBN flakes deposited on an array of Al$_2$O$_3$ pillars. The laser excitation wavelength was 532 nm and the emission was dispersed by a 1200 g/mm diffraction grating aligned to a thermoelectrically-cooled CCD detector array. Any pixels with a total intensity significantly greater than the background intensity were identified as potential hBN colour centres, with the average number of such pixels being approximately one per pillar. A representative selection of such pixels is indicated by red circles and are numbered. Although the majority of such pixels are found at the pillar edges, some are indeed found near the centre of the pillars (eg. Point 2). **(c)** Corresponding PL spectra for the selection of points labelled in Fig. 3b. A theoretical fit to each spectrum (red lines) was performed using a linear combination of Lorentzian and Gaussian functions. The PL peaks at 586 nm (635 and 650 nm) for Point 1, 579 nm (622 and 632 nm) for Point 2, 551 nm (606 nm) for Point 3, and 579 nm for Point 4, are assigned to the ZPL (PSB) of hBN defect colour centres, which were fit using Lorentzian (Gaussian) functions. The PSB is on average observed to be ∼ 53 nm (170 meV) red-shifted relative to the ZPL, in agreement with the literature[22]. The ubiquitous sharp peaks at 580 nm and 620 nm correspond to the G and 2D Raman peaks of graphene, respectively. The spectrum for Point 2 also shows a sharp peak at 574.7 nm, which corresponds to the Raman peak for the $E_{2g}$ mode in hBN and was fit using a Lorentzian function.

Having fabricated the devices for the localisation of hBN colour centre emission, we then performed confocal photoluminescence (PL) mapping to determine whether the localisation of emission to the Al$_2$O$_3$ pillars had been achieved. Confocal PL mapping was performed using a laser scanning confocal microscope (Olympus FluoView FV1200) equipped with a 473 nm continuous-wave (cw) excitation laser and a photomultiplier tube detector for measuring emission in the 560–660 nm wavelength range using a band pass filter. Figure 3a shows the result of a confocal PL map of an area of the sample where hBN flakes have been deposited and where 6 × 6 arrays of 40-nm-thick 4.75 $\mu$m × 4.75 $\mu$m Al$_2$O$_3$ pillars have been patterned on graphene. This pillar lateral dimension was chosen since it had been found to result in the localisation of approximately one colour centre emitter on average per pillar. As can be seen in Fig. 3a, emission is clearly localised within areas of the Al$_2$O$_3$ pillars, as it is virtually absent within areas where Al$_2$O$_3$ had been etched down to the graphene layer. It is also observed that emission is more intense towards the edges of pillars, which is consistent with SEM images that showed an accumulation of hBN flakes towards the edges of pillars (Fig. 2d).

In order to confirm that localised emission observed in Fig. 3a is due to hBN defect colour centres, we obtained hyperspectral PL maps of the pillar arrays using a confocal Raman microscope operating in PL mode (Renishaw InVia) equipped with a 532 nm cw excitation laser and a 1200 g/mm diffraction grating aligned to a thermoelectrically-cooled charge-coupled device (CCD) detector array. Figure 3b shows a map of the total intensity at each pixel of the hyperspectral map and is consistent with the confocal map in Fig. 3a. Any pixels in Fig. 3b with an intensity significantly greater than the background intensity were identified as potential hBN colour centres. While the average number of such pixels was found to be approximately one per pillar, a representative selection of such pixels is indicated in Fig. 3b by red circles and are numbered, with their corresponding PL spectrum shown in Fig. 3c. Following a baseline correction using the asymmetrically reweighted penalised least squares smoothing (ARPLS) scheme[26,27], a theoretical fit to each PL spectrum (red lines in Fig. 3c) was performed using a linear combination of Lorentzian and Gaussian functions. Given the typical wavelengths and linewidths of the zero-phonon line (ZPL) emission from hBN colour centres reported in the literature[28] and the typical red-shifts of the phonon sideband (PSB) relative to the ZPL reported in the literature[22], we assign the PL peaks at 586 nm (635 and 650 nm) for Point 1, 579 nm (622 and 632 nm) for Point 2, 551 nm (606 nm) for Point 3, and 579 nm for Point 4, to the ZPL (PSB) of hBN defect colour centres, which were fit using Lorentzian (Gaus-sian) functions. Meanwhile, the ubiquitous sharp peaks at 580 nm and 620 nm correspond to the *G* and 2*D* Raman peaks of graphene, respectively, thereby confirming the preservation of the underlying graphene within the device structure. The spectrum for Point 2 also shows a sharp peak at 574.7 nm, which corresponds to the Raman peak for the $E_{2g}$ mode in hBN[29] and was fit using a Lorentzian function. It is worth noting that the differing ZPL emission wavelengths in Fig. 3c are likely due to variations in the induced strain fields on hBN defect colour centres[30,31], as a result of various factors such as the lattice mismatch between hBN and the substrate material. Additionally, variations in the linewidth of ZPL emission are likely due to variations in the thickness of hBN flakes hosting defect colour centres[3].

It is important to note that, aside from pixels assigned to hBN colour centres, the faint background intensity observed on the pillars (especially towards the edges) in Figs. 3a and 3b do not show any remarkable features in their spectra and is attributed to the Raman signal of hBN and scattering effects from the pillars. Away from the pillars, no distinct features were observed in the spectra, aside from the Raman signals of hBN and graphene. However, at some locations in the confocal maps of Figs. 3a and 3b, emission can be seen away from the pillars. Although this is relatively insignificant compared to emission at the pillar sites, we attribute this to incomplete etching of the $Al_2O_3$ layer, leaving residual islands that are sufficiently thick such that colour centres in hBN flakes are protected from quenching. As mentioned earlier, these islands are the result of the polycrystalline nature of the ALD film, leading to a non-uniform etch rate of the $Al_2O_3$ layer. Therefore, the etch time will need to be further optimised to ensure maximal removal of these residual $Al_2O_3$ islands away from the pillars, while also minimising undercutting of the $Al_2O_3$ pillars.

We have demonstrated a novel approach to the localisation of hBN colour centre emission by using patterned dielectric layers grown on graphene via ALD. While colour centre emission was quenched within areas where hBN was deposited directly on graphene due to charge transfer, it was maintained where hBN was deposited on micron-sized $Al_2O_3$ pillars which served as barriers to charge transfer. Crucially, our approach allows for device architectures where graphene layers are used as top and bottom electrodes for the application of vertical electric fields. Moreover, the ability to use ALD to grow the $Al_2O_3$ layers directly on graphene allows for the control of tunnel barrier thicknesses with atomic layer precision. Beyond the present work, the degree of localisation may be enhanced by a combination of increasing the density of colour centres[3,32] and reducing the lateral dimensions of the pillars. Nevertheless, the results of our work open the way for the electrical control of deterministically localised hBN colour centres and the development of a wide range of scalable device technologies based on hBN defect colour centres.

The authors gratefully acknowledge funding from the Sir Henry Royce Institute (Cambridge Royce facilities grant EP/P024947/1 and Sir Henry Royce Institute - recurrent grant EP/R00661X/1). The authors also acknowledge Ermanno Miele and Jonathan Griffiths for assistance with the electron-beam lithography process.